\documentclass[suppldata]{interact}

\usepackage{epstopdf}
\usepackage[caption=false]{subfig}

\usepackage{url}
\usepackage{multirow}

\theoremstyle{plain}

\theoremstyle{definition}

\theoremstyle{remark}

\RequirePackage{amsthm,amsmath,amsfonts,amssymb,bbold}
\usepackage{algorithm}
\usepackage{algpseudocode}

\usepackage{color}

\newcommand{\spacer}[1]{{{\hspace{5mm}{#1}\hspace{5mm}}}}
\newcommand{\vspacer}[1]{{{\vspace{2mm}}}}

\begin{document}


\title{Standardization of Weighted Ranking Correlation Coefficients}

\author{
\name{P.~Lombardo\textsuperscript{a}}
\affil{\textsuperscript{a}Data and Information Department, Eutelsat, France}
}

\maketitle

\begin{abstract}
A fundamental problem in statistics is measuring the correlation between two rankings of a set of items. 
Kendall's $\tau$ and Spearman's $\rho$ are well established correlation coefficients whose symmetric structure guarantees
 zero expected value between two rankings randomly chosen with uniform probability.

In many modern applications, however, greater importance is assigned to top-ranked items, motivating weighted variants of these coefficients.
Such weighting schemes generally break the symmetry of the original formulations, resulting in a non-zero expected value under independence and compromising the interpretation of zero correlation.
We propose a general standardization function $g(\cdot)$ that transforms a ranking correlation  coefficient $\Gamma$ into a standardized form $g(\Gamma)$ with zero expected value under randomness. 
The transformation preserves the domain $[-1,1]$, satisfies the boundary conditions, is continuous and increasing, and reduces to the identity for coefficients that already satisfy the zero-expected-value property.

The construction of $g(x)$ depends on three distributional parameters of $\Gamma$: its mean, variance, and left variance; since their exact calculation becomes infeasible for large ranking lengths $n$, we develop accurate numerical estimates based on Monte Carlo sampling combined with polynomial regression to capture their dependence on $n$.

\end{abstract}


\section{Introduction}

An important problem in statistics is defining the correlation between two rankings of a set of items; this problem has gathered an increasing amount of attention in recent decades because rankings are ubiquitous in many fields, including 
search engines,
recommendation systems,
information retrieval, 
machine learning model comparison,
evaluation of natural language processing systems, and
applied sciences
\cite{croux2010influence, de2016comparing, yannakakis2015ratings, lombardo2020top}.

Ranking correlation also provides a non-parametric (distribution-free) alternative to more traditional correlation coefficients such as Pearson's $r$, which captures linear relationships.
This alternative is particularly useful in cases when variable distributions are skewed or heavy-tailed: in these cases, rank-based measures have been shown to have lower variability and to produce lower deviation between an estimate over a sample and the value on the whole population compared with the Pearson coefficient.
For example, \cite{de2016comparing} show that for heavy-tailed variables Spearman's $\rho$ typically has lower variability than Pearson's $r$ and its estimation on a subsample often corresponds more accurately to the Pearson correlation coefficient of the whole population than the Pearson estimate itself.

Moreover, when assessing monotonic relationships, including non-linear ones, and without assuming a specific distributional form of the variables, ranking correlation coefficients are often more appropriate than measures of linear correlation such as Pearson's r \cite{hauke2011comparison, de2016comparing, puth2015effective}.

Another reason for using rankings instead of 
the original variables is to mitigate inconsistencies that frequently emerge when comparing numeric variables originating from heterogeneous sources
\cite{ammar2011ranking,negahban2017rank,yannakakis2015ratings}. 

The problem of rank correlation
was originally addressed through two closely related but distinct approaches by Kendall and Spearman, who introduced respectively Kendall $\tau$ \cite{kendall1938new}
and Spearman $\rho$ \cite{spearman1961proof}.

Subsequently, several alternative definitions have been proposed; in particular, considerable attention has been devoted to weighted ranking correlation coefficients designed to reflect the greater importance of top-ranked items. 
In many contemporary applications, such as recommender systems, search engines, and semantic matching, users are primarily exposed to the first few positions of a ranking.
As a consequence, discrepancies occurring at the top of the list have a substantially larger practical impact than discrepancies occurring at lower ranks. Evaluation measures that treat all positions uniformly may therefore fail to capture relevant performance differences \cite{de2008integrating,de2015semantics,lops2011content,mladenic1999text,giunchiglia2004s,li2014semantic,wan2016deep,blest2000theory,pinto2005weighted,dancelli2013two,iman1987measure,maturi2008new,shieh1998weighted,vigna2015weighted,webber2010similarity}.

Different extensions of Spearman's $\rho$ and Kendall's $\tau$ have been proposed to incorporate the greater importance of top ranks
\cite{pinto2005weighted,dancelli2013two,vigna2015weighted,lombardo2020top}, with weighting schemes tailored to specific application needs.

The original (unweighted) Spearman and Kendall coefficients are characterized by a symmetry: for any pair of rankings yielding a given coefficient value, there exists another pair yielding the opposite value; 
as a consequence, when averaging  over all possible pairs of rankings, the expected value of the coefficient is zero.
This symmetry property does not generally hold for weighted ranking correlation coefficients. The introduction of position-dependent weights breaks the invariance underlying the original formulations, and the expected value of the coefficient under random permutations may differ from zero. As a result, zero no longer represents a natural benchmark corresponding to absence of correlation. This loss of interpretability complicates empirical comparisons and may lead to misleading conclusions, particularly when weighted coefficients are used for model evaluation.

Despite the growing use of weighted ranking correlation coefficients, the problem of restoring a zero expected value under random permutations has not been addressed in a general and systematic manner.
For this reason, we propose a standardization framework that transforms a ranking correlation coefficient into a form with zero expected value when averaged over all possible rankings pairs, while preserving its structural properties.

In Section \ref{sec:summary_of_coeffs} we summarize the main standard and weighted ranking correlation coefficients, focusing on their expected value under independence.
In Section \ref{sec:standardized_coeff} we introduce the standardization function that maps a generic ranking coefficient into a standardized version with zero expected value as well as procedures to estimate the distributional parameters required for the construction of the transformation.
In Section \ref{sec:example_corr} we present practical examples in the context of movie recommendation, illustrating the differences between standard and weighted coefficients and highlighting the effect of the proposed standardization.
Finally, in Section \ref{sec:results}, we report the numerical results of the standardization procedure, verifying that the transformed coefficients satisfy the required properties.

\section{Summary of Ranking Correlation Coefficients}
\label{sec:summary_of_coeffs}

In this section, we briefly summarize the most relevant ranking correlation coefficients, focusing on rankings without ties.

\subsection{Standard Ranking Correlation Coefficients}
\label{sec:standard_coeff}

To formulate a unified treatment of the main ranking correlation coefficients, it is convenient to start from the general form introduced by Kendall \cite{kendall1938new}: 
\begin{equation}\label{eq:gamma}
    \Gamma = \frac{\sum_{i,j}a_{ij}b_{ij}}{\sqrt{\sum_{i,j}a_{ij}^2 \sum_{i,j}b_{ij}^2}},
\end{equation}
where $a_{ij}$ ($b_{ij}$) is a matrix that depends on the first (second) ranking to be compared, with indices $i,j$ running over all items. This formulation
contains both Spearman's $\rho$ and Kendall's $\tau$.

Indeed, setting $a_{ij}=a_j-a_i$ and $b_{ij} = b_j-b_i$, where $\{a_i\}$ and $\{ b_i\}$ denote the rankings to be compared, yields Spearman's $\rho$ \cite{spearman1961proof}
\begin{equation}\label{eq:spearman}
    \rho = \frac{\sum_{i,j}(a_i-a_j)(b_i-b_j)}{\sqrt{\sum_{i,j}(a_i-a_j)^2\sum_{i,j}(b_i-b_j)^2}}.
\end{equation}
Introducing the quantities $\bar a = \sum_ja_j$ and $\bar b = \sum_jb_j$, Eq.~\ref{eq:spearman} can be reformulated as the Pearson correlation between the rank variables, which reduces the computation to sums over a single index, with a significant decrease in the computation complexity, i.e.,
\begin{equation}
    \rho = \frac{\sum_{i}(a_i-\bar a)(b_i-\bar b)}{\sqrt{\sum_i(a_i-\bar a)^2\sum_i(b_i-\bar b)^2}}.
\end{equation}

On the other hand, Kendall's $\tau$ \cite{kendall1938new,kruskal1958ordinal} is obtained from Eq.~\ref{eq:gamma} by setting $a_{ij}={\mathrm{sgn}}(a_j-a_i)$ and $b_{ij} = {\mathrm{sgn}}(b_j-b_i)$: 
\begin{equation}\label{eq:kendall}
    \tau = \frac{\sum_{i,j}{\mathrm{sgn}}(a_j-a_i)\,{\mathrm{sgn}}(b_j-b_i)}{n(n-1)},
\end{equation}
where $n$ is the length of the rankings and
\begin{eqnarray}
\nonumber  {\mathrm{sgn}(x)}   = & 1 &{\mathrm{if}}\  x>0, \\
         {\mathrm{sgn}(x)}   = & 0 &{\mathrm{if}}\ x=0, \\
\nonumber {\mathrm{sgn}(x)}   = & -1 &{\mathrm{if}}\ x<0.
\end{eqnarray}
In the absence of ties, Eq.~\ref{eq:kendall} can be rewritten in the more familiar form 
\begin{equation}
    \tau = \frac{N_{\mathcal{C}(\{a_i\},\{b_i\})}-N_{\mathcal{D}(\{a_i\},\{b_i\})}}{n(n-1)},
\end{equation}
where $\mathcal{C}(\{a_i\},\{b_i\})$ ($\mathcal{D}(\{a_i\},\{b_i\})$) denotes the set of concordant (discordant) pairs of distinct elements , i.e., the pairs $i,j$ with $i\neq j$ for which $a_j-a_i$ and $b_j-b_i$ have the same (different) sign, while $N_{\mathcal{C}(\{a_i\},\{b_i\})}$ ($N_{\mathcal{D}(\{a_i\},\{b_i\})}$) represents the cardinality of such set, 
and $n(n-1)$ is the total number of ordered pairs.

\subsection{Weighted Ranking Correlation Coefficients}
\label{sec:w_coeff}

Several extensions of Spearman's $\rho$ and Kendall's $\tau$ have been proposed to address use cases in which top ranks are more important than lower ranks. 
Most of these metrics can be reduced to the general form in Eq.~\ref{eq:gamma}
\cite{lombardo2020top}.

In particular, defining $a_{ij} = \sqrt{w_iw_j}\,(a_j-a_i)$ and $b_{ij} = \sqrt{w_iw_j}\,(b_j-b_i)$,  where $w_i$ is the normalized weight associated to the $i$-th position in the rankings, yields the weighted Spearman coefficient, which can be written as
\begin{equation}\label{eq:w_spearman}
         \rho_{\mathrm w} = \frac{\sum_i w_i(a_i-\bar a)(b_i-\bar b)}{\sqrt{\sum_i w_i(a_i^2-\bar a^2) \sum_i w_i(b_i^2-\bar b^2)}},
\end{equation}
where $\bar a=\sum_i w_i a_i$ and $\bar b=\sum_i w_i b_i$.

On the other hand, defining $a_{ij} = \sqrt{w_iw_j}\,{\mathrm{sgn}}(a_j-a_i)$ and $b_{ij} = \sqrt{w_iw_j}\,{\mathrm{sgn}}(b_j-b_i)$, leads to the weighted Kendall coefficient
\begin{equation}\label{eq:w_kendall}
     \tau_{\mathrm w} = \frac{\sum_{i\neq j} w_i w_j\,{\mathrm{sgn}}(a_j-a_i) \,{\mathrm{sgn}}(b_j-b_i)}{\sum_{i\neq j}w_i w_j},
\end{equation}
or equivalently
\begin{equation}\label{eq:w_kendall_C} 
     \tau_{\mathrm w} = \frac{\sum_{i,j\in \mathcal{C}(\{a_i\}, \{b_i\})}w_iw_j-\sum_{i,j\in \mathcal{D}(\{a_i\}, \{b_i\})}w_iw_j}{\sum_{i\neq j}w_iw_j}.
\end{equation}

The coefficients in Eqs.~\ref{eq:w_spearman} and \ref{eq:w_kendall} encompass, up to notational differences, many of the weighted ranking coefficients proposed in the literature to take into account the greater importance of the top ranks
\cite{pinto2005weighted,dancelli2013two,vigna2015weighted,lombardo2020top}.

In the literature, different weighting schemes are used to assign the importance to the $i$ index combining the ranks $a_i$ and $b_i$; they are typically based on single-rank weights $f(a_i)$ and $f(b_i)$, where $f$ is a monotonically decreasing function of the index $i$. A paradigmatic example is harmonic weighting,
\begin{equation}\label{eq:f_harmonic}
    f(i) = \frac{1}{i},
\end{equation}
characterized by the divergence of the normalization factor $\sum_{i=1}^{n}f(i)$ when the length $n$ of the ranking tends to infinity. Thus, in the infinite-ranking limit, no single index $i$ provides a finite relative contribution, which may be undesirable in some applications.

Another commonly used weighting function, characterized by the convergence of the normalization factor, is the inverse quadratic
\begin{equation}\label{eq:f_quadratic}
    f(i)=\frac{1}{(i+n_0)^2},
\end{equation}
where $n_0$ is a parameter controlling the relative contribution of the first rank(s) \cite{lombardo2020top}.

Since Eqs. \ref{eq:w_spearman}, \ref{eq:w_kendall}, and \ref{eq:w_kendall_C} involve pairs of ranks $a_i,b_j$, it is necessary to specify how the weights $f(a_i)$ and $f(b_j)$ derived from the two rankings are combined. The two most common approaches are 
the additive combination
\cite{dancelli2013two,vigna2015weighted,lombardo2020top}
\begin{equation}\label{eq:a_weights}
    w_i=\frac{f(a_i)+f(b_i)}{2\sum_jf(j)}
\end{equation}
and the multiplicative combination \cite{dancelli2013two,vigna2015weighted}
\begin{equation}\label{eq:m_weights}
    w_i=\frac{f(a_i)f(b_i)}{\sum_jf(a_j)f(b_j)}.
\end{equation}

For large $n$, the multiplicative scheme cannot discriminate different rankings when they differ only by the exchange of a top rank $a_i$ and a low rank $b_i$, as the factor $f(b_i)$ makes the quantity $w_i$ negligible. For this reason, the additive scheme is often preferable in applications where such discrimination is relevant \cite{dancelli2013two,lombardo2020top}.

\subsection{Ranking--Permutation Duality}
\label{sec:duality}

To simplify the calculation of the expected value over all ranking pairs, it is worth noting that, in the absence of ties, the set of rankings of length $n$ is in bijection with the group $\mathbb{\Pi}_n$ of permutations on $\mathbb{Z}_n=\{0, 1,..,n -1 \}$.
Therefore, choosing two rankings $\{a_i\}$ and $\{b_i\}$ 
is equivalent to choosing two permutations $\pi_a$ and $\pi_b$ with $a_i=\pi_a(i)$ and $b_i=\pi_b(i)$; averaging over all pairs of rankings is thus equivalent to averaging over all pairs of permutations.

Thanks to this correspondence, the coefficient $\Gamma$ in Eq.~\ref{eq:gamma} can be written as a function of the two permutations $\pi_a$ and $\pi_b$ introduced above. 
Moreover, due to the identity $\sum_{i,j}h(\pi_a(i), \pi_b(j))=\sum_{i,j}h([\pi_a^{-1} \circ \pi_a](i), [\pi_a^{-1} \circ \pi_b](j))=\sum_{i,j}h(i, [\pi_a^{-1}\circ\pi_b](j))$, where $\circ$ represents composition and $h(i,j)$ is a generic function $\mathbb{Z}_n^2\to \mathbb{R}$,
we can write $\Gamma$ as a function of a single permutation $\pi$, with $\pi = \pi_a^{-1} \circ \pi_b$.
Another way to state the same property is that for any pair of rankings $\{a_i\}$, $\{b_i\}$,
the ranking correlation $\Gamma(\{a_i\}, \{b_i\})$ can be written as the correlation between the natural-sort ranking $\{1,2, .., n\}$ and $\{\pi_a^{-1}(b_i)\}$.

For future convenience, we rewrite here the coefficients introduced in Sections \ref{sec:standard_coeff} and \ref{sec:w_coeff} using the permutations formalism.
The standard Spearman $\rho$ and Kendall $\tau$ correlation coefficients become, respectively,
\begin{equation}\label{eq:spearman_sigma}
     \rho(\pi) =
    \frac{\sum_{i,j}(j-i)[\pi(j)-\pi(i)]}{\sqrt{\sum_{i,j}(j-i)^2\sum_{i,j}[\pi(j)-\pi(i)]^2}} 
\end{equation}
and
\begin{equation}\label{eq:kendall_sigma}
    \tau(\pi) = \frac{N_{\mathcal{C}(\pi)}-N_{\mathcal{D}(\pi)}}{n(n-1)},
\end{equation}
where $\mathcal{C}(\pi)$ ($\mathcal{D}(\pi)$) denotes the set of ordered pairs $i,j$ with $i\neq j$ and for which the natural sorting and $\pi$ are concordant (discordant), i.e., ${\mathrm{sgn}}(j-i) = {\mathrm{sgn}}(\pi(j)-\pi(i))$ (${\mathrm{sgn}}(j-i) \neq {\mathrm{sgn}}(\pi(j)-\pi(i))$).

Similarly, the weighted Spearman and Kendall coefficients in Eqs.~\ref{eq:w_spearman} and \ref{eq:w_kendall} can be rewritten as
\begin{equation}\label{eq:w_spearman_sigma}
     \rho_{\mathrm w}(\pi) =
    \frac{\sum_{i\neq j}w_iw_j(j-i)[\pi(j)-\pi(i)]}{\sqrt{\sum_{i\neq j}w_iw_j(j-i)^2\sum_{i\neq j}w_iw_j[\pi(j)-\pi(i)]^2}}
\end{equation}
and
\begin{equation}\label{eq:w_kendall_sigma}
     \tau_{\mathrm w}(\pi) =
    \frac{\sum_{i, j\in \mathcal{C}(\pi)}w_iw_j-\sum_{i, j\in \mathcal{D}(\pi)}w_iw_j}{\sum_{i\neq j}w_iw_j}.
\end{equation}

As discussed in Section \ref{sec:w_coeff}, Eq.~\ref{eq:w_spearman_sigma} can be rewritten in terms of a single sum:
\begin{equation}\label{eq:w_spearman_sigma_singlesum}
    \rho_{\mathrm w}(\pi)=\frac{\sum_i[i-\langle i\rangle][\pi(i)-\langle\pi(i)\rangle]}{\sqrt{\sum_i[i-\langle i\rangle]^2\,\sum_i[\pi(i)-\langle\pi(i)\rangle]^2}},
\end{equation}
where $\langle e(i)\rangle = \sum_iw_ie(i)$ denotes the weighted average.
For this reason, the (weighted) Spearman coefficient is computationally more efficient to evaluate than the (weighted) Kendall coefficient.

\subsection{Expected Value for Ranking Correlation Coefficients}

Given a ranking correlation coefficient $\Gamma$ depending on two rankings of length $n$, 
we can express its expected value over all ranking pairs using the permutation formalism introduced in Section \ref{sec:duality} 
\begin{equation}\label{eq:gamma_ev_sigma}
E(\Gamma) = \frac{1}{n!}\sum_{\pi \in\mathbb{\Pi}_n}\Gamma(\pi),
\end{equation}
where $\mathbb{\Pi}_n$ is the set of the $n!$ permutations of $\mathbb{Z}_n$.

Since two independently and uniformly chosen rankings are expected to be uncorrelated on average%
\footnote{Rankings are drawn uniformly at random from the set of all permutations.},
a desirable property of a ranking correlation coefficient is $E(\Gamma)=0$.
The standard Spearman $\rho$ and Kendall $\tau$ satisfy this property.
However, the general form in Eq.~\ref{eq:gamma} does not necessarily have zero expectation, and, in particular,  the weighted coefficients in Eqs.~\ref{eq:w_spearman} and \ref{eq:w_kendall} do not guarantee this property.

\subsubsection{Expected Value for Standard Ranking Correlation Coefficients.}
\label{sec:EV_standard}

For both the standard Spearman's $\rho$ and Kendall's $\tau$, the expected value in Eq.~\ref{eq:gamma_ev_sigma} can be computed using symmetry arguments.

For every permutation $\pi$, consider the permutation $\tilde{\pi}$ defined by $\tilde{\pi}(i)=n+1-\pi(i)$. One immediately verifies that 
$\rho(\tilde \pi)=-\rho(\pi)$ and similarly $\tau(\tilde \pi)=-\tau(\pi)$.
Since permutations can be paired in this way, their contributions cancel in the sum over $\mathbb{\Pi}_n$, yielding
$E(\rho)=E(\tau)=0$.

\subsubsection{Expected Value for Weighted Ranking Correlation Coefficients.}

A key difference between Eqs.~\ref{eq:spearman_sigma}-\ref{eq:kendall_sigma} and their weighted counterparts is the presence of weights $w_i$, which may depend on $\pi$ (e.g., under additive or multiplicative schemes).
For this reason, the symmetry argument used in Section \ref{sec:EV_standard} no longer applies, and cancellations are not guaranteed.

Equation \ref{eq:gamma_ev_sigma} can therefore be used to calculate the expected value of any weighted ranking coefficient introduced in Sections \ref{sec:standard_coeff} and \ref{sec:w_coeff}.
Thanks to the ranking--permutation duality, the computation requires summing over $n!$ permutations rather than $(n!)^2$ ranking pairs.
Nevertheless, the exact calculation becomes infeasible for large $n$ and for this reason, in Section \ref{sec:distr_par_estimation_summary}, we introduce a procedure to estimate $E(\Gamma)$.

\section{Standardization Procedure}
\label{sec:standardized_coeff}

In this section, we describe a general procedure for constructing a standardized version of any ranking correlation coefficient $\Gamma(\pi)$ that can be written in the form of Eq.~\ref{eq:gamma}, including those defined in Eqs.~\ref{eq:spearman_sigma}, \ref{eq:kendall_sigma}, \ref{eq:w_spearman_sigma}, and \ref{eq:w_kendall_sigma}.
For this purpose, we introduce a standardization function $g(x)$ that transforms the coefficient while preserving its structural properties, so that the transformed coefficient $g(\Gamma)$ has zero expected value under independence. The goal is to restore the interpretability property enjoyed by classical symmetric coefficients, without altering the ordinal information encoded by $\Gamma$.

To simplify the notation, we introduce the following quantities
\begin{eqnarray}\label{eq:notation_x}
    \bar\Gamma &=& E(\Gamma)\\
    \nonumber p(\gamma) &=& \frac{1}{n!}\sum_{\pi\in \mathbb{\Pi}_n}\delta(\gamma-\Gamma(\pi)),
\end{eqnarray}
where $\delta(x)$ denotes the Dirac delta distribution, so that $p(\gamma)$ represents the distribution of $\Gamma$ over the interval $[-1,1]$.

The function $g(x)$ depends on $p(\gamma)$ --- which in turn depends on the type of coefficient (Spearman or Kendall), the weighting scheme (additive or multiplicative), the weighting function $f(i)$, and the ranking length $n$ --- 
and maps $\Gamma$ into a quantity with zero expected value. 
\begin{itemize}
    \item[0.  ] The expected value $E(g(\Gamma))$ over all possible pairs of rankings is zero, i.e., $\int_{-1}^{1}{\mathrm{d}}\gamma\, p(\gamma)\, g(\gamma)=0$.
\end{itemize}
On the other hand, since we want the shifted coefficient $g(\Gamma)$ to maintain the mathematical properties of the original coefficient $\Gamma$, $g(x)$ must also satisfy the following consistency conditions.
\begin{enumerate}
    \item Same domain: $g(x)$ maps the interval $[-1,1]$ into itself, i.e., $-1\leq g(x)\leq 1$ for any $x\in [-1,1]$.
    \item Boundary conditions: $g(-1)=-1$ and $g(1)=1$.
    \item Continuity of $g(x)$ and its first derivative in the interval $[-1,1]$.
    \item Monotonicity: $g(x)$ is increasing, i.e., $g(x_2)> g(x_1)$ for any $x_2>x_1$ in the interval $[-1,1]$. 
    \item No shift for standard coefficients: in the case of a correlation coefficient characterized by a symmetric distribution $p(\gamma)=p(-\gamma)$, such as the standard Spearman $\rho$ and Kendall $\tau$, $g$ reduces to the identity function $g(x)=x$.
\end{enumerate}
The monotonicity requirement (condition 4) ensures the consistency of the rankings $\Gamma$ and $g(\Gamma)$, i.e.,  given any pair of permutations $\pi_1$ and $\pi_2$, if $\Gamma(\pi_1)>\Gamma(\pi_2)$, then $g(\Gamma(\pi_1)) > g(\Gamma(\pi_2))$.
Note that, since the original coefficient $\Gamma$ is already constrained in the $[-1,1]$ domain, consistency condition 1 is a consequence of conditions 2 and 4.

We search for a piecewise polynomial function in the domains $[-1,\bar\Gamma]$ and $[\bar\Gamma,1]$,
\begin{eqnarray}\label{eq:gamma_shifted_general}
g(x) &=& \sum_{a=0}^{\alpha}g_a(x-\bar\Gamma)^a,\ {\mathrm{if}}\ x <\bar\Gamma\\
\nonumber g(x) &=& \sum_{a=0}^{\alpha}h_a(x-\bar\Gamma)^a,\ {\mathrm{if}}\ x \geq\bar\Gamma,
\end{eqnarray}
where $\{g_a\}$ and $\{h_a\}$ are real parameters constrained by the consistency condition.

A linear function ($\alpha=1$) does not allow all consistency conditions to be satisfied simultaneously; therefore, the quadratic case ($\alpha=2$) is the simplest admissible choice and
%
Eq.~\ref{eq:gamma_shifted_general} reduces to
\begin{eqnarray}\label{eq:gamma_shifted}
g(x) &=& g_0+g_1(x-\bar\Gamma)+g_2(x-\bar\Gamma)^2,\ {\mathrm{if}}\ x <\bar\Gamma\\
\nonumber g(x) &=& g_0+g_1(x-\bar\Gamma)+h_2(x-\bar\Gamma)^2,\ {\mathrm{if}}\ x \geq\bar\Gamma,
\end{eqnarray}
where we set $h_0=g_0$ and $h_1=g_1$ to enforce the continuity of $g(x)$ and $g'(x)$ (consistency condition 3).
The boundary conditions (consistency condition 2) determine other two constraints on the parameters, namely
\begin{eqnarray}\label{eq:coeff_g2_h2}
g_2 &=& -\frac{1+g_0}{(1+\bar\Gamma)^2}+\frac{g_1}{1+\bar\Gamma},\\
\nonumber h_2 &=& \frac{1-g_0}{(1-\bar\Gamma)^2}-\frac{g_1}{1-\bar\Gamma}.
\end{eqnarray}

To enforce the property of zero expected value (condition 0), we introduce two additional quantities that, together with the expected value $\bar\Gamma$, characterize the distribution $p(\gamma)$, namely, the variance 
\begin{equation}\label{eq:variance}
 V=\int_{-1}^{1}{\mathrm{d}}\gamma\, p(\gamma)\,(\gamma-\bar\Gamma)^2
\end{equation}
and the left variance, defined as
\begin{equation}\label{eq:variance_left}
 V^{\mathrm{\ell}}=\int_{-1}^{\bar\Gamma}{\mathrm{d}}\gamma\, p(\gamma)\,(\gamma-\bar\Gamma)^2.
\end{equation}
The quantity $V^{\mathrm{\ell}}$ represents the contribution to the total variance coming from values of $\gamma$ below the mean, and therefore captures the asymmetric allocation of dispersion around $\bar\Gamma$.
To avoid division by zero in the derivation of the standardization parameters, a special treatment is required when
\begin{equation}\label{eq:flat_variance_ratio}
 \frac{V^{\mathrm{\ell}}}{V} = \frac{1+\bar\Gamma}{2},
\end{equation}
a condition that we refer to as the \emph{flat variance ratio} and includes the symmetric distribution ($V^{\mathrm{\ell}}=V/2$, $\bar\Gamma=0$) as a special case.

In the presence of numerical estimates of $\bar\Gamma$, $V$, and $V^{\mathrm{\ell}}$, we relax condition \eqref{eq:flat_variance_ratio} to
 \begin{equation}\label{eq:flat_variance_ratio_loose}
 \left|\frac{V^{\mathrm{\ell}}}{V} - \frac{1+\bar\Gamma}{2}\right| <\varepsilon_{\mathrm{f}},
\end{equation}
where $\varepsilon_{\mathrm{f}}\ll 1$ accounts for numerical approximation errors.

\subsection{Zero-Expected-Value and Monotonicity Requirements for Flat Variance Ratio}
\label{sec:flat_variance_ratio}

In the flat variance ratio case, the zero-expected-value requirement (condition 0) uniquely determines $g_0$ as
\begin{equation}\label{eq:cond3_Azero}
g_0=-\frac{V\, \bar\Gamma}{1-V-\bar\Gamma^2}.
\end{equation}

The only free parameter remaining in Eq.~\ref{eq:gamma_shifted} is therefore $g_1$. 
The monotonicity requirement (consistency condition 4) yields the constraint
\begin{equation}\label{eq:cond5_Azero}
    0\leq g_1\leq \frac{2\,{\mathrm{min}}(1-\bar\Gamma-V,1+\bar\Gamma-V)}{1-\bar\Gamma^2-V}.
\end{equation}
Although the denominator in \eqref{eq:cond5_Azero} 
is strictly positive, the numerator may in principle be non-positive for arbitrary distributions 
 $p(\gamma)$.  However, for all weighting schemes considered in this work that produce a flat variance ratio, the upper bound in \eqref{eq:cond5_Azero} is strictly positive, ensuring the existence of admissible values of $g_1$. 

Since any value of $g_1$ satisfying \eqref{eq:cond5_Azero} guarantees both zero expectation and monotonicity, the transformation is not uniquely determined. To obtain a canonical choice and preserve comparability across weighting schemes, we set $g_1=1$ whenever this value satisfies \eqref{eq:cond5_Azero}; otherwise we set $g_1$ equal to the upper bound in \eqref{eq:cond5_Azero}.
The symmetric case $\bar\Gamma=0$ corresponds to classical Spearman and Kendall coefficients. In this case, Eq.~\ref{eq:cond3_Azero} gives $g_0=0$, and Eq.~\ref{eq:cond5_Azero} reduces to $0\leq g_1\leq 2$. Choosing $g_1=1$ yields $g_2=h_2=0$ and therefore $g(x)=x$, as required by consistency condition 5.

\subsection{Zero-Expected-Value Requirement for non-Flat Variance Ratio}
\label{sec:non_flat_variance_ratio_zev}

In this case, the zero-expected-value requirement (condition 0) leads to the relation
\begin{equation}\label{eq:cond3_Anonzero}
g_1=\frac{(1-\bar\Gamma^2)V-4\bar\Gamma V^{\mathrm{\ell}}-(1-\bar\Gamma^2)^2}{(1-\bar\Gamma^2)[2V^{\mathrm{\ell}}-V(1+\bar\Gamma)]}g_0
 + \frac{2(1+\bar\Gamma^2)V^{\mathrm{\ell}}-(1-\bar\Gamma^2)V}{(1-\bar\Gamma^2)[2V^{\mathrm{\ell}}-V(1+\bar\Gamma)]},
\end{equation}
where $g_0$ is the only remaining free parameter in Eq.~\ref{eq:gamma_shifted};
since it represents the transformed value of the mean ($g_0 = g(\bar\Gamma)$), it must lie in the interval $[-1,1]$.

The monotonicity requirement (consistency condition 4) then induces a set of inequalities restricting the admissible values of $g_0$. 
For several combinations of coefficient type, weighting scheme, weighting function, and ranking lengths, these constraints are compatible with $g_0=0$, which constitutes a natural choice, as it implies $g(\bar\Gamma)=0$.

However, for some combinations of the aforementioned quantities, the choice $g_0=0$ violates the monotonicity condition.
In the following section, we introduce a general algorithmic procedure to determine an admissible value of $g_0$ that satisfies all consistency requirements.

\subsection{Monotonicity Requirement for non-Flat Variance Ratio}
\label{sec:non_flat_variance_ratio_monotonic}

The monotonicity requirement (consistency condition 4) is equivalent to condition $g'(x)> 0$ almost everywhere, except possibly at isolated points where $g'(x)=0$.
Since $g'(x)$ is piecewise linear in $x$ and $g(x)$ is not constant, this condition is in turn equivalent  to requiring
(\emph{i}) $g'(-1) \geq 0$, (\emph{ii}) $g'(\bar\Gamma)\geq 0$, and (\emph{iii}) $g'(1)\geq 0$, which correspond, respectively, to 
\begin{eqnarray}
\label{eq:g_derivate>0}
  &(\emph{i})&\left[2(1-\bar\Gamma)-B\right] g_0 \geq C-2(1-\bar\Gamma),\\
   \nonumber &(\emph{ii})&B g_0 \leq - C,\ {\mathrm{and}}\\
  \nonumber &(\emph{iii})&\left[2(1+\bar\Gamma)+B\right] g_0 \leq 2(1+\bar\Gamma)-C,
\end{eqnarray}
where
\begin{eqnarray}
    &&B=\frac{(1-\bar\Gamma^2)V-4\bar\Gamma V^{\mathrm{\ell}} - (1-\bar\Gamma^2)^2}{2V^{\mathrm{\ell}} - V(1 + \bar\Gamma)},\\
     \nonumber &&C = \frac{2(1 + \bar\Gamma^2)V^{\mathrm{\ell}} - (1-\bar\Gamma^2) V}{2V^{\mathrm{\ell}} - V(1 + \bar\Gamma)}.
\end{eqnarray}

Inequalities \ref{eq:g_derivate>0} define the admissible bounds for $g_0$. 
Similarly to the case of the flat variance ratio, the existence of a solution is not guaranteed for an arbitrary probability distribution $p(\gamma)$.
However, for each specific $p(\gamma)$ (i.e., for each tuple of values $\bar\Gamma$, $V$, and $V^{\mathrm{\ell}}$) associated with the weighting schemes described in this paper, we find that there always exists a value of $g_0$ satisfying all inequalities \ref{eq:g_derivate>0}.

Appendix \ref{sec:g0_determination_algorithm} provides a general procedure to set $g_0$ based on the values of $\bar\Gamma$, $V$, and $V^{\mathrm{\ell}}$, ensuring that inequalities \ref{eq:g_derivate>0} are satisfied whenever a feasible solution exists, while raising an error in case the solution does not exist.

The high-level logic is the following.
By default, we assign $g_0=0$, which corresponds to mapping the expected value of $\Gamma$ to 0; if the inequalities are not satisfied, we assign to $g_0$ the admissible value that satisfies all the bounds and is closest to $0$. In case the bounds are not compatible or require $g_0$ outside the domain $[-1,1]$, a \emph{bound consistency error} is raised. 
We did not encounter bound consistency errors for any of the weighting schemes analyzed in this paper.

To account for numerical errors in the estimations of the quantities $\bar\Gamma$, $V$, and $V^{\mathrm\ell}$, each condition of the form \emph{if} $\alpha(\bar\Gamma, V, V^{\rm\ell}) = 0$ is relaxed to \emph{if} $|\alpha(\bar\Gamma, V, V^{\rm\ell})|<\varepsilon_{\mathrm{b}}$, with $\varepsilon_{\mathrm{b}}\ll 1$ (in our experiments, $\varepsilon_{\mathrm{b}}=10^{-8}$).

We conclude that the function $g(x)$ defined in Eq.~\ref{eq:gamma_shifted} is completely determined once the quantities $\bar\Gamma$, $V$, and $V^{\mathrm{\ell}}$ characterizing the distribution $p(\gamma)$ are known.

\subsection{Distribution Parameters Estimation Summary}
\label{sec:distr_par_estimation_summary}

The parameters $\bar\Gamma$, $V$, and $V^{\mathrm{\ell}}$ can be  computed exactly for small ranking lengths (e.g., for $n\lesssim 10$), while for larger ranking lengths, they  are estimated via Monte Carlo sampling over the permutation space, followed by polynomial regression to model their dependence on $n$.
The details of the estimation procedure for the mean $\bar\Gamma$ and the variances $V$ and $V^{\mathrm{\ell}}$ are provided in Appendices \ref{sec:gamma_bar_estimate} and \ref{sec:var_estimation}, respectively.

\begin{figure}[ht]
    \centering
    \includegraphics[width=\columnwidth]{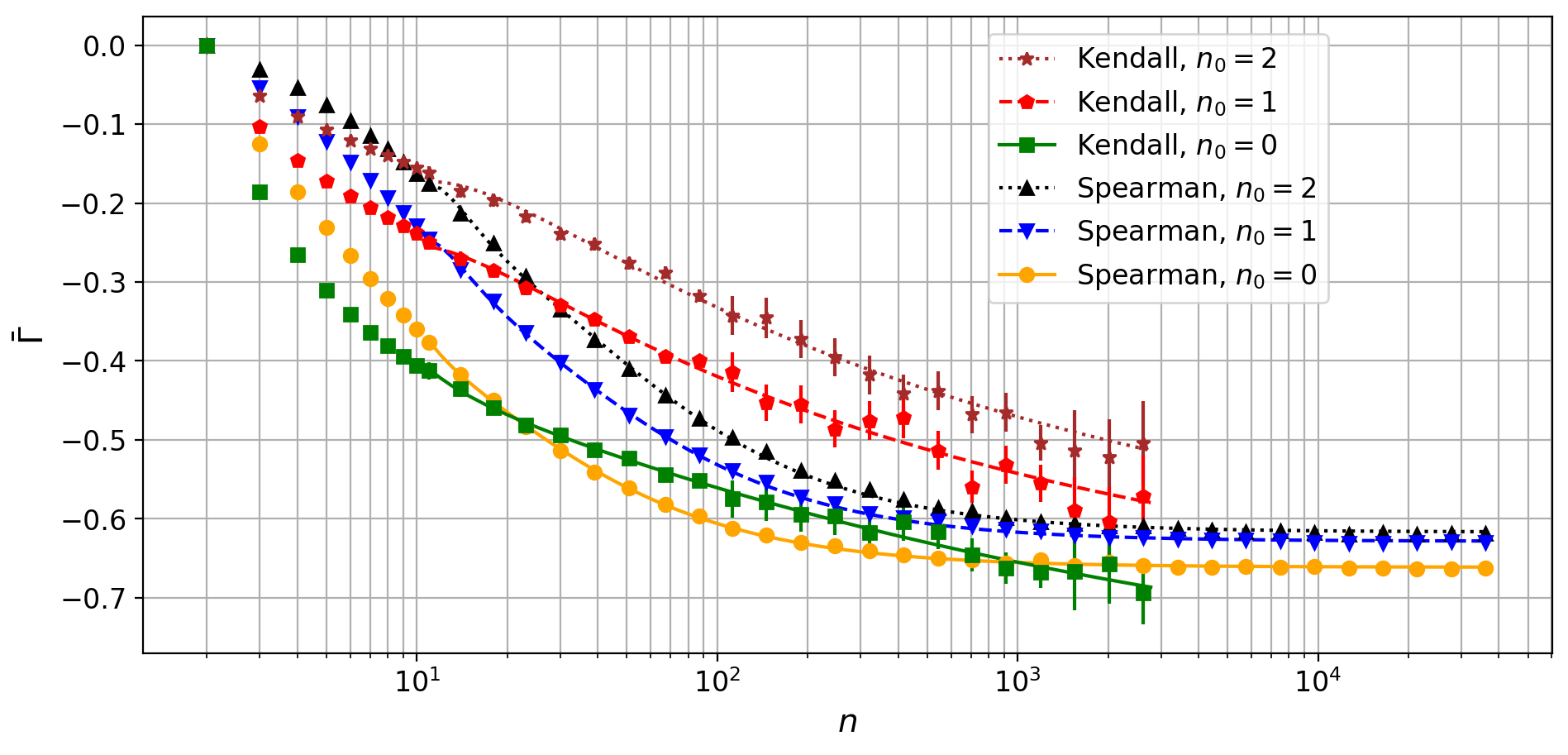}
    \caption{Results of the $\bar\Gamma$ estimation procedure described in Section \ref{sec:gamma_bar_estimate} for Kendall with multiplicative weight and Spearman with additive weight, in both cases with weighting function $f(i)=1/(i+n_0)^2$.}
    \label{fig:gamma_bar_plot}
\end{figure}

Figure \ref{fig:gamma_bar_plot} shows illustrative examples of estimation of $\bar\Gamma$, namely, Kendall with multiplicative weight and Spearman with additive weight, both using the weighting function $f(i)=1/(i+n_0)^2$. Additional details and complete estimation results are reported in Appendix \ref{sec:estimation_results}.
%

\subsection{Operational Summary of the Standardization Procedure}

The proposed standardization of the weighted Spearman $\rho$ and Kendall $\tau$ coefficients can be practically implemented using the python function \emph{standard\_gamma\_calc()} available at \url{https://github.com/plombardML/ranking_correlation}.

The steps to replicate the behaviour of the python function are the following:
\begin{itemize}
  \item Estimate $\bar\Gamma$, $V$, and $V^{\mathrm{\ell}}$
  \begin{itemize}
    \item If $n \leq n_{\mathrm{exact}} = 10$, use the exact values reported in Appendix \ref{sec:estimation_results}.
    \item If $n > n_{\mathrm{exact}}$, estimate the parameters using the regression coefficients reported in Appendix \ref{sec:estimation_results}, applying the transformation $x(n)$ defined in Eqs.~\ref{eq:n_to_x} and \ref{eq:n_to_x_log} when required.
  \end{itemize}
    \item Verify the variance ratio condition (Eq.~\ref{eq:flat_variance_ratio} or inequality~\ref{eq:flat_variance_ratio_loose}) and determine $g_0$ and $g_1$ according to the flat or non-flat regime.
    \item Compute $g_2$ and $h_2$ using Eqs.~\ref{eq:coeff_g2_h2}.
    \item Define the standardization function $g(x)$ via Eq.~\ref{eq:gamma_shifted}.
    \item Compute the standardized coefficient as $g(\Gamma)$, where $\Gamma$ denotes the original (unstandardized) coefficient.
\end{itemize}

\section{Application to Movie Recommendation}\label{sec:example_corr}

To illustrate the differences between standard and weighted ranking correlation, and to highlight the effect of the proposed standardization, we compare a ground truth ranking with several alternative rankings in the context of movie recommendation%
\footnote{The code to replicate this example is available in the python notebook \emph{movie\_recommendatation\_application} at \url{https://github.com/plombardML/ranking_correlation}.}.

We use the Movielens 100k dataset%
\footnote{Dataset available at \url{https://grouplens.org/datasets/movielens/100k/}},
which consists of integers ratings between 1 and 5; 
we split the dataset into two subsets according to the timestamp. 
Subset \emph{A} contains ratings collected between \emph{1997-09-20} and \emph{1998-03-07}, corresponding approximately to $80\%$ of the entries in the dataset, while subset \emph{B} contains the remaining $20\%$, collected between \emph{1998-03-08} and \emph{1998-04-23}.
In many practical applications, user feedback is simplified to binary responses (e.g., \emph{OK} or \emph{KO}), in order to minimize the user effort and increase the response rate. While this approach is operationally simpler, it may lead to information loss compared to a more structured feedback like the 1-5 rating. To investigate this effect, we construct for each rating a binary version where a value of 1 (OK) is assigned to original ratings equal to  4 or 5, and 0 (KO) otherwise%
\footnote{The threshold 4 is commonly used in practice to distinguish positive from non-positive feedback in 1-5 ratings.}.

We define as ground truth the ranking based on the average original rating computed on subset A. This ranking is then compared with the following alternatives:
\begin{enumerate}
  \item random sorting (baseline)
  \item ranking based on the simplified rating on subset B
  \item ranking based on the simplified rating on subset A
  \item a perturbed version of the ground truth ranking in which the last movie is moved to the first position and all remaining movies are shifted down by one position.
\end{enumerate}
The last construction is intentionally artificial and is introduced to illustrate the sensitivity of weighted correlation coefficients to severe errors occurring at the top of the ranking.
These rankings are shown in Table \ref{tab:movies} for the top 10 movies according to the ground truth (original rating on subset A).
%
%
\begin{table}[ht]
\tbl{Top 10 movies according to the ground truth (rate A) and their rankings according to the ranking methods described in section \ref{sec:example_corr}.}
{\begin{tabular}{cccccc}
\toprule
Title  & Rate A & Random & Simple rate B & Simple rate A & Last first \\ \midrule
Maya Lin: A Strong Clear Vision (1994) & 1      & 310    & 603                                                     & 1                                                       & 1382                                                 \\
Prefontaine (1997)                     & 2      & 742    & 1                                                       & 2                                                       & 1                                                    \\
Star Kid (1997)                        & 3      & 266    & 2                                                       & 3                                                       & 2                                                    \\
Ayn Rand: A Sense of Life (1997)       & 4      & 824    & 604                                                     & 4                                                       & 3                                                    \\
Golden Earrings (1947)                 & 5      & 779    & 1112                                                    & 5                                                       & 4                                                    \\
Close Shave, A (1995)                  & 6      & 661    & 159                                                     & 21                                                      & 5                                                    \\
Wrong Trousers, The (1993)             & 7      & 77     & 152                                                     & 22                                                      & 6                                                    \\
Horseman on the Roof, The (1995)       & 8      & 185    & 854                                                     & 6                                                       & 7                                                    \\
Perfect Candidate, A (1996)            & 9      & 746    & 605                                                     & 7                                                       & 8                                                    \\
Pather Panchali (1955)                 & 10     & 487    & 3                                                       & 8                                                       & 9                                                    \\ \bottomrule
\end{tabular}}
\label{tab:movies}
\end{table}
%
%

For a recommendation system, the highest focus is on the top ranks: an error in the early positions is much more relevant than an error in the lower ranks. For this reason, we compare the rankings 1-4 with the ground truth using a weighted correlation coefficient.

\begin{table}[ht]
\tbl{Comparison of the 4 movie rankings described in Section \ref{sec:example_corr} with the ground truth. The correlation is calculated using  standard Spearman $\rho$ and Kendall $\tau$ and their weighted version $\rho_{\mathrm w}$ and $\tau_{\mathrm w}$ with additive weighting scheme and the parameters indicated, both in the standardized and non-standardized version.}
{\begin{tabular}{ccc|cccc} \toprule
Coefficient                              & $f(i)$                       & Standardized & Random    & Simple rate B & Simple rate A & Last first \\ \midrule
Spearman                               &                              &        & $2.7\%$   & $60.7\%$                                                & $96.0\%$                                              & $99.6\%$                                                                 \\ \midrule
\multirow{4}{*}{Weighted Spearman} & \multirow{2}{*}{$1/i$}       & no     & $-33.1\%$ & $36.6\%$                                                & $97.4\%$                                              & $59.9\%$                                                                 \\
                                    &                              & yes    & $0.9\%$   & $50.1\%$                                                & $97.9\%$                                              & $67.8\%$                                                                 \\ \cmidrule{2-7} 
                                    & \multirow{2}{*}{$1/(i+1)^2$} & no     & $-71.5\%$ & $-8.9\%$                                                & $85.1\%$                                              & $-1.6\%$                                                                 \\
                                    &                              & yes    & $-14.1\%$ & $55.2\%$                                                & $99.2\%$                                              & $61.0\%$                                                                 \\ \midrule
Kendall                              &                              &        & $1.8\%$   & $48.1\%$                                                & $84.0\%$                                              & $99.7\%$                                                                 \\ \midrule
\multirow{4}{*}{Weighted Kendall} & \multirow{2}{*}{$1/i$}       & no     & $-27.5\%$ & $31.0\%$                                                & $87.8\%$                                              & $75.5\%$                                                                 \\
                                    &                              & yes    & $0.1\%$   & $38.6\%$                                                & $87.9\%$                                              & $76.2\%$                                                                 \\ \cmidrule{2-7} 
                                    & \multirow{2}{*}{$1/(i+1)^2$} & no     & $-52.6\%$ & $16.1\%$                                                & $96.3\%$                                              & $26.6\%$                                                                 \\
                                    &                              & yes    & $1.5\%$   & $70.2\%$                                                & $99.9\%$                                              & $77.2\%$                                                            \\    \bottomrule
\end{tabular}}
\label{tab:example_corr}
\end{table}

The correlation results are reported in Table \ref{tab:example_corr}. Several aspects 
deserve attention.
(i) Without standardization, weighted coefficients may indicate strong negative correlation even for the random ranking, which should represent the absence of correlation. This illustrates the interpretability issue caused by the non-zero expectation of weighted coefficients.
(ii) The ranking based on simplified ratings from subset A (the same subset used to construct the ground truth) is strongly correlated with the ground truth according to all coefficients; this suggests that, in this case, binary feedback would have been a solid alternative with minimal loss of information. 
(iii) The ranking based on the simplified rating from subset B may appear weakly correlated or even negatively correlated with the ground truth when the non-standardized weighted coefficients are used; after standardization, however, the coefficients provide a more coherent and interpretable assessment.
Finally, (iv) the \emph{last-first} perturbation achieves correlation values above $99.5\%$ according to the standard Spearman and Kendall coefficients, while the weighted coefficients reveal a substantial degradation in agreement.
This behaviour is consistent with recommendation settings, where a severe error in the highest-ranked positions may cause loss of trust in the recommender and has a much stronger impact than errors occurring lower in the list (such as those in the ranking based on the simplified rating in subset B). In such contexts, weighted and properly standardized coefficients provide a more meaningful assessment of ranking quality.

This example illustrates that standardization restores the interpretation of zero correlation as absence of correlation in the weighted setting.

\section{Results and Conclusion}
\label{sec:results}

In this work, we have introduced a general procedure to standardize ranking correlation coefficients $\Gamma$ that can be expressed in the form of Eq.~\ref{eq:gamma}. 
The proposed construction preserves coefficients that already satisfy the zero-expected-value property under independence, such as Spearman’s $\rho$ and Kendall’s $\tau$ (Eqs.~\ref{eq:spearman_sigma}, \ref{eq:kendall_sigma}), while providing a principled correction for weighted variants (Eqs.~\ref{eq:w_kendall_sigma}, \ref{eq:w_spearman_sigma_singlesum}), in which this property no longer holds.
The  procedure consists of transforming the coefficient according to a function $g(x)$ that preserves the main structural properties of $\Gamma$ while enforcing zero expected value under independence; 
in this way, the standardized coefficient retains the original ordering information while restoring the interpretation of zero correlation as statistical independence, even in the weighted setting.

The construction of $g(x)$ depends on three distributional parameters of $\Gamma$: its mean, variance, and left variance; 
the exact calculation of these parameters for a given ranking length $n$ requires summing over $(n!)^2$ ranking pairs or, exploiting the ranking--permutation duality discussed in  Section \ref{sec:duality}, summing over $n!$  permutations.
Such computations rapidly become unfeasible for large $n$; for this reason, we propose accurate numerical estimates based on Monte Carlo sampling combined with polynomial regression to model the $n$ dependence, which allow us to approximate the distributional parameters efficiently while maintaining high accuracy.

\begin{figure}[ht]
    \centering
    \includegraphics[width=\columnwidth]{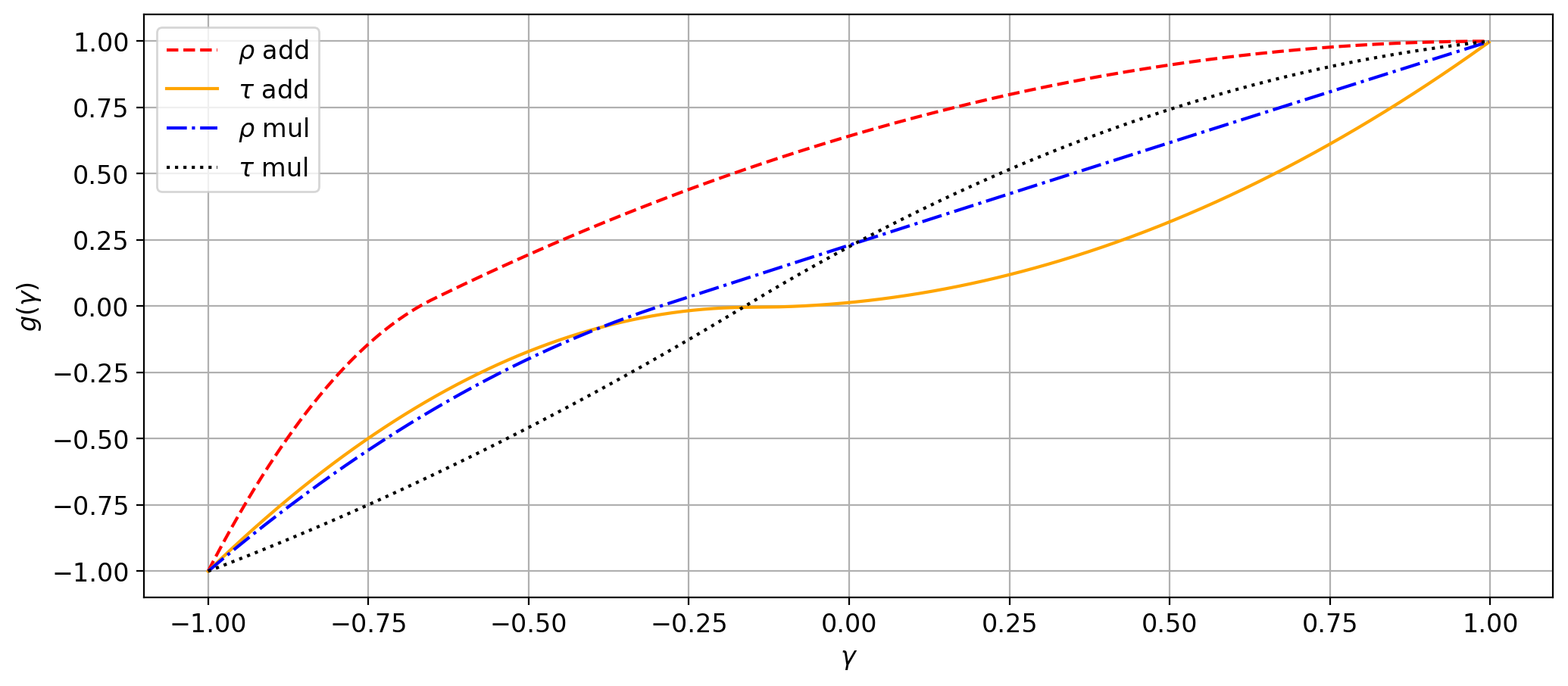}
    \caption{Standardization function $g(x)$ for the Spearman $\rho$ (Kendall $\tau$) coefficient with additive and multiplicative weights, with $f(i)=1/i^2$ ($f(i)=1/i$) and ranking length $n=500$ ($n=30$).}
    \label{fig:result_g}
\end{figure}
In Figure \ref{fig:result_g} we report the standardization function $g(x)$ for representative examples, namely the weighted Spearman coefficient with additive (red dashed) and multiplicative (blue dot-dash) weights, with $f(i)=1/i^2$ and ranking length $n=500$ and the weighted Kendall coefficient with additive (yellow solid) and multiplicative (black dotted) weights, with $f(i)=1/i$ and $n=30$.

\begin{figure}[ht]
    \centering
    \includegraphics[width=\columnwidth]{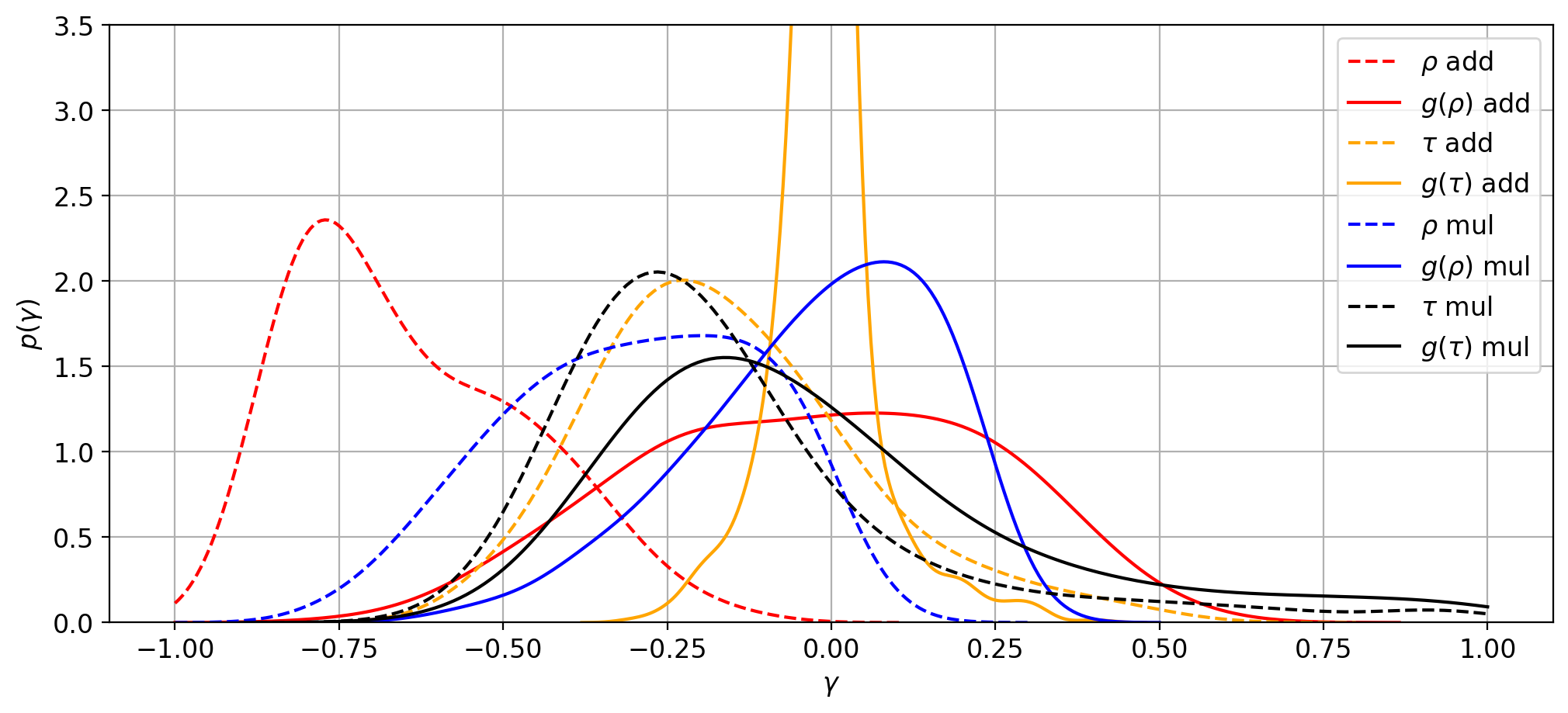}
    \caption{Estimate of the distribution function $p(\gamma)$ before and after the standardization of the coefficients for the Spearman $\rho$ (Kendall $\tau$) coefficient with additive and multiplicative weights, with $f(i)=1/i^2$ ($f(i)=1/i$) and ranking length $n=500$ ($n=30$).}
    \label{fig:result_p}
\end{figure}
Figure \ref{fig:result_p} shows an estimate of the distribution $p(\gamma)$ before (dashed line) and after (solid line) the standardization of the coefficients for the same examples described above. The distributions are estimated from samples of 10,000 points with kernel density estimation; the bandwidth is selected according to Scott's rule and scaled by a factor 0.4.

The results confirm that the standardization function $g(x)$ satisfies the properties described in Section \ref{sec:standardized_coeff}, namely, it maps the domain $[-1,1]$ into itself, respects the boundary conditions, is monotonic increasing, and the standardized distribution $p(g(\gamma))$ is coherent with the zero-expected-value requirement.

In Section \ref{sec:example_corr} we presented practical examples in the context of movie recommendation systems, where errors in the highest-ranked positions have a substantially stronger impact than errors occurring lower in the list.
These examples demonstrate that weighted and properly standardized coefficients provide a more meaningful assessment of ranking quality in such contexts.
Moreover, they illustrate that standardization restores the interpretation of zero correlation as absence of correlation in the weighted setting, as shown by the empirical distributions in Figure~\ref{fig:result_p}.

The proposed framework provides a principled and general solution to the bias induced by weighting schemes, enabling meaningful comparison across ranking lengths and weighting strategies.

While the estimation of the distributional parameters relies on Monte Carlo sampling combined with polynomial regression, the computational feasibility depends on both the coefficient and the adopted weighting scheme. 
For weighted Spearman coefficients, the procedure can be applied to very large ranking lengths (up to $n=40{,}000$ in the present analysis) and, for certain weighting schemes, the empirical behaviour of the fitted parameters clearly indicates convergence toward a stable asymptotic regime as $n\to\infty$. 
For weighted Kendall coefficients, the combinatorial structure results in more restrictive computational bounds (up to $n=3{,}000$ in our experiments), although similar asymptotic trends are observed for specific classes of weights.

Future work may aim at establishing analytical asymptotic expressions for the distributional parameters, deriving theoretical guarantees for the observed convergence, and quantifying the approximation error introduced by the Monte Carlo and regression-based estimation procedure.

\bibliographystyle{plain}
\bibliography{RankCorrBib}

\appendix

\section{Determination of $g_0$}
\label{sec:g0_determination_algorithm}


\begin{algorithm}[ht]
\caption{Setting $g_0$}\label{g_0_setting}

\textbf{Input:} 
$\bar\Gamma$, $V$, $V^{\mathrm{\ell}}$

\vspacer

\textbf{Initialize:}
$A =2 V^{\mathrm{\ell}} - V  (1 + \bar\Gamma)$,
$B = \frac{(1 - \bar\Gamma^2) V - 4 \bar\Gamma\,V^{\mathrm{\ell}} - (1 - \bar\Gamma^2)^2}{A}$,
$C =\frac{2 (1 + \bar\Gamma^2)V^{\mathrm{\ell}} - (1 - \bar\gamma^2) V}{A}$,
$D = \frac{C-2(1-\bar\Gamma)}{2(1-\bar\Gamma)-B}$,
$E = -\frac{C}{B}$, 
$F =  \frac{2(1+\bar\Gamma)-C}{2(1+\bar\Gamma)+B}$

\vspacer

Set
$b_1^{\mathrm{low}}=-1$,
 $b_1^{\mathrm{up}}=1$ 
\Comment{Bound \emph{i}}

\textbf{if} $2(1-\bar\Gamma) - B >0$ 
\textbf{then} $b_1^{\mathrm{low}}=D$

\textbf{else} 

\quad\textbf{if} $2(1-\bar\Gamma) - B <0$
\textbf{then} $b_1^{\mathrm{up}}=D$

\quad\textbf{else}

\quad\quad\textbf{if} $|C - 2 (1 - \bar\Gamma)| > \varepsilon_{\mathrm{b}}$
\textbf{then}   \textbf{Bound Consistency Error}

\quad\quad \textbf{end if}

\quad \textbf{end if}

\textbf{end if}

\vspacer

Set $b_2^{\mathrm{low}}=-1$, $b_2^{\mathrm{up}}=1$
\Comment{Bound \emph{ii}}

\textbf{if} $B >0$
\textbf{then} $b_2^{\mathrm{low}}=E$

\textbf{else}

\quad\textbf{if} $B <0$
\textbf{then} $b_2^{\mathrm{up}}=E$

\quad\textbf{else}

\quad\quad\textbf{if} $|C| > \varepsilon_{\mathrm{b}}$  
\textbf{then} \textbf{Bound Consistency Error}

\quad\quad \textbf{end if}

\quad \textbf{end if}

\textbf{end if}

\vspacer

Set $b_3^{\mathrm{low}}=-1$, $b_3^{\mathrm{up}}=1$
\Comment{Bound \emph{iii}}

\textbf{if} $2(1+\bar\Gamma) + B >0$
\textbf{then} $b_3^{\mathrm{up}}=F$

\textbf{else}

\quad\textbf{if} $2(1+\bar\Gamma) + B<0$
\textbf{then} $b_3^{\mathrm{low}}=F$

\quad\textbf{else}

\quad\quad\textbf{if} $|C - 2 (1 + \bar\Gamma)| > \varepsilon_{\mathrm{b}}$
\textbf{then} \textbf{Bound Consistency Error}

\quad\quad \textbf{end if}

\quad \textbf{end if}

\textbf{end if}

\vspacer

Set $b^{\mathrm{low}} = \max(b_1^{\mathrm{low}}, b_2^{\mathrm{low}}, b_3^{\mathrm{low}})$, $b^{\mathrm{up}} = \min(b_1^{\mathrm{up}}, b_2^{\mathrm{up}}, b_3^{\mathrm{up}})$
\Comment{Bounds combination}

\textbf{if} $b^{\mathrm{up}}<b^{\mathrm{low}}$
\textbf{then} \textbf{Bound Consistency Error}

\textbf{end if}

Set $g_0=0$

\textbf{if} $b^{\mathrm{low}}>g_0$
\textbf{then} $g_0=b^{\mathrm{low}}$

\textbf{else}

\quad\textbf{if} $b^{\mathrm{up}}<g_0$
\textbf{then} $g_0=b^{\mathrm{up}}$

\quad\textbf{end if}

\textbf{end if}

\vspacer

\textbf{return} $g_0$

\end{algorithm}

Algorithm \ref{g_0_setting} provides the pseudo-code of the procedure used to determine an admissible value of $g_0$ from the distributional parameters $\bar\Gamma$, $V$, and $V^{\mathrm{\ell}}$, as described in Section \ref{sec:non_flat_variance_ratio_monotonic}. 
The algorithm returns a value of $g_0$ that satisfies the monotonicity constraints \eqref{eq:g_derivate>0}, whenever such a solution exists; otherwise, it signals that no admissible value can be found.

\section{Estimation of the Expected Value $\bar\Gamma$}
\label{sec:gamma_bar_estimate}

We describe here a procedure to obtain a numerical estimate of the expected value in Eq.~\ref{eq:gamma_ev_sigma} for a wide range of ranking lengths $n$. The approach consists of:
\begin{itemize}
    \item Computing the exact value $\bar\Gamma$     for $n\leq n_{\mathrm{exact}}$, where we set $n_{\mathrm{exact}}= 10$.
    \item Selecting a set of ranking lengths $S_n$ for each of which a numerical estimate of $\bar\Gamma$ is obtained via numerical sampling.
    \item Fitting  the numerical estimates with a polynomial regression model to approximate $\bar\Gamma$ as a function of the ranking length $n$ over a broader domain.
\end{itemize}

\subsection{Numerical Sampling}\label{sec:numerical_sampling}

A standard approach to estimate $\bar\Gamma$ defined in Eq.~\ref{eq:gamma_ev_sigma} is Monte Carlo sampling, that is, instead of computing the average on the complete set of permutations $\mathbb{\Pi}_n$, we calculate the average on a set of samples $S_{\pi}^n=\{\pi^{(n)}_1, \pi^{(n)}_2, \ldots, \pi^{(n)}_{N_{\mathrm{samp}}}\}$ containing $N_{\mathrm{samp}}$ permutations, independently drawn (with the possibility of repetition) from $\mathbb{\Pi}_n$ with uniform probability.

The coefficient $\Gamma$ maps the set of permutations $\mathbb{\Pi}_n$ into a set of $n!$ values $\gamma=\Gamma(\pi)$ in the domain $[-1,1]$, with distribution $p(\gamma)$ characterized by mean $\bar\Gamma$ and variance $V$.
Accordingly, $\Gamma$ maps $S_{\pi}^n$ into a sample $G_n=\{ \gamma_1, .. \gamma_{N_{\mathrm{samp}}}\}$, where the $\{\gamma_a\}$ are independent and identically distributed random variables with distribution $p(\gamma)$.
By the law of large numbers, the sample mean  $\bar\gamma=\sum_{a}\gamma_a/N_{\mathrm{samp}}$ converges to  $\bar\Gamma$ as $N_{\mathrm{samp}}$ increases.

An estimate of the variance of $\bar\gamma$ is $\mathrm{V}(\bar\gamma) = \frac{V}{N_{\mathrm{samp}}}$,
where the variance $V$ is estimated from the sample as $\frac{1}{N_{\mathrm{samp}} - 1} \sum_{a=1}^{N_{\mathrm{samp}}} (\gamma_a - \bar\gamma)^2$.

\subsection{Polynomial Regression}
\label{sec:polynomial_reg}

To extend the numerical sampling results $\{\Gamma_n\}$ to other ranking lengths $n$, we perform a polynomial regression. 
More in detail, we fit a polynomial of order $D_{\gamma}$ 
\begin{equation}
y(x)=\sum_{d=0}^{D_{\gamma}} y_d\, x^d,
\end{equation}
where $x=x(n)$ is a monotonic transformation of the ranking length $n$ and the coefficients $y_d$ are obtained by minimizing the weighted sum of squared residuals 
\begin{equation}\label{eq:regression_sr}
    SSR = \sum_{n \in S_{\mathrm{train}}}\frac{1}{V(\bar\gamma_n)}\left[ y(x(n)) - \bar\gamma_n\right]^2,
\end{equation}
using inverse-variance weighting.

\subsubsection{Training set.}

The numerical sampling is performed on a set $S_{\mathrm{train}}$ that contains ranking lengths $n$ spanning the range to which we are particularly interested. Since this range is wide, we choose logarithmically distributed ranking lengths $n_a$, i.e., 
$S_{\mathrm{train}}=\{ n_a = [q^a]\}$, where $a$ is an integer with $a_{\mathrm{min}}\leq a \leq a_{\mathrm{max}}$ and $[x]$ represents the rounding of $x$ to the closest integer.
In this work, we chose the following values.
\begin{itemize}
  \item $q=1.3$, so that we have approximately 9 points $n_a$ before an increase of a factor 10.
  \item $a_{\mathrm{min}}=9$, corresponding to $n_{a_{\mathrm{min}}}=11$, as the exact value $\bar\Gamma$ can be easily calculated up to $n=10$.
  \item $a_{\mathrm{max}}$ is 30 (40) for Kendall (Spearman) coefficient, corresponding to a maximum value of 2620 (36119) for $n_a$%
\footnote{\label{foot:kendal_vs_spearman_training}The difference in the training set for Spearman and Kendall coefficients is due to the different computational complexity highlighted in Section \ref{sec:duality}.}%
.    
\end{itemize}

As described in Section \ref{sec:numerical_sampling}, for each of the ranking lengths $n$ in $S_{\mathrm{train}}$, we calculate $\bar\gamma$ based on a sample of $N_{\mathrm{samp}}(n)$ permutations, which has the following values%
\footnote{See footnote \ref{foot:kendal_vs_spearman_training}.}%
.
\begin{itemize}
    \item Spearman coefficient: $N_{\mathrm{samp}}(n)$ is $10^5$ if $n\leq 100$ and $10^4$ if $n>100$.
    \item Kendall coefficient: $N_{\mathrm{samp}}(n)$ is $10^4$ if $n\leq 100$, $10^3$ if $100<n\leq 1500$, and 200 if $n>1500$.
\end{itemize}

\subsubsection{Choice of x(n).}

In this work, we consider two different monotonic transformations $x(n)$, namely
\begin{equation}\label{eq:n_to_x}
    x(n) = \frac{1}{n}
\end{equation}
and
\begin{equation}\label{eq:n_to_x_log}
    x(n) = \frac{1}{{\log}(n)}.
\end{equation}
In both cases, the limit $x\to 0$ corresponds to the limit $n\to\infty$, and thus the intercept $y_0$ represents the estimate of the value of $\bar\Gamma$ in the large $n$ limit. 
Since $n = 1/x$ in Eq.~\ref{eq:n_to_x} and $n = \exp(1/x)$ in Eq.~\ref{eq:n_to_x_log}, the mapping in Eq.~\ref{eq:n_to_x} yields a more gradual growth of $n$ as $x \to 0$. 
This typically improves numerical stability and extrapolation accuracy in the large-$n$ regime, making Eq.~\ref{eq:n_to_x} preferable when the empirical behavior of $\bar\gamma$ supports it.
\begin{figure}[ht]
    \centering
    \includegraphics[width=\columnwidth]{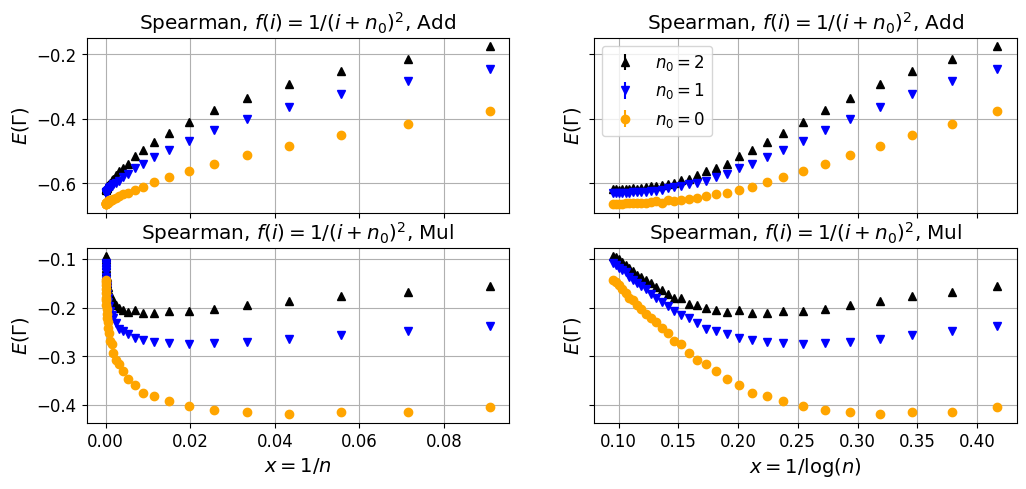}
    \caption{Numerical estimate of $\bar\Gamma$ as a function of $x$ defined in Eq.~\ref{eq:n_to_x} (left) and Eq.~\ref{eq:n_to_x_log} (right) for Spearman coefficient with additive (top) and multiplicative (bottom) weight $w_i=1/(i+n_0)^2$.}
    \label{fig:log_2_spearman}
\end{figure}

In Figure \ref{fig:log_2_spearman} we show the plot of the sampling estimates of $\bar\gamma$ as a function of $x$ for the Spearman coefficient with weighting function $f(i)=1/(i+n_0)^2$, comparing additive (top) and multiplicative (bottom) weighting schemes and the $x(n)$ transformations described in Eqs.~\ref{eq:n_to_x} (left) and \ref{eq:n_to_x_log} (right).
We may notice that, while for the additive weighting scheme both transformations $x(n)$ are acceptable, for the multiplicative weighting scheme Eq.~\ref{eq:n_to_x} determines a large variation of $\bar\gamma$ associated with a small variation of $x$ as $x\to 0$, while Eq.~\ref{eq:n_to_x_log} provides an almost linear dependence of $\bar\gamma(x)$, making it the preferable choice to perform a regression.
With analogous considerations, for Kendall coefficient and weighting function $f(i)=1/(i+n_0)^2$ we choose Eq.~\ref{eq:n_to_x} (\ref{eq:n_to_x_log}) for the additive (multiplicative) weighting scheme, while for both Spearman and Kendall coefficients and weighting function $f(i)=1/i$ we choose Eq.~\ref{eq:n_to_x_log}.

\subsubsection{Choice of the polynomial degree $D_{\gamma}$.}
\label{sssec:degree}

The degree $D_{\gamma}$ of the polynomial is chosen taking into account the Mean Square Error (MSE) between the numerical estimate $\bar\gamma_n$ and the regression estimate $y(x(n))$ but also the complexity of the model.
\begin{figure}[ht]
    \centering
    \includegraphics[width=\columnwidth]{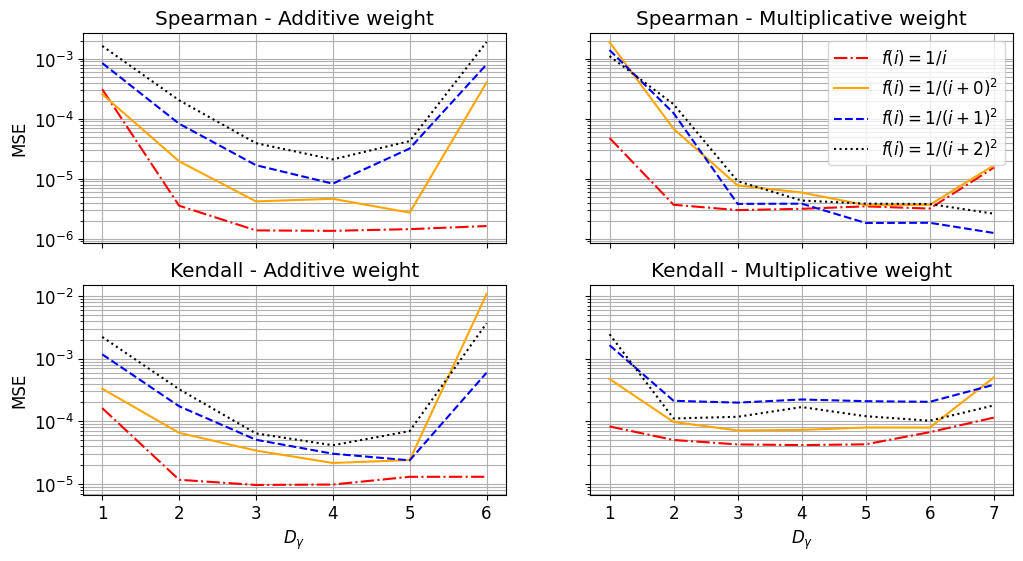}
    \caption{MSE as a function of the polynomial degree $D_{\gamma}$ for the Spearman and Kendall coefficients with the different weighting schemes described in Section \ref{sec:w_coeff}.}
    \label{fig:deg_optimization}
\end{figure}
In order to avoid overfitting, we increase the degree $D_{\gamma}$ only in case of substantial improvement in terms of MSE.

More in detail, we calculate the MSE for several polynomial degrees $D_{\gamma}$, as shown in Figure \ref{fig:deg_optimization}, considering the ratio of the MSEs associated with subsequent degrees. We allow increasing the degree to $D_{\gamma}$ only if the ratio ${\mathrm{MSE}}(D_{\gamma})/{\mathrm{MSE}}(D_{\gamma}-1)$ is below $0.8$ or if ${\mathrm{MSE}}(D_{\gamma}+1)/{\mathrm{MSE}}(D_{\gamma}-1)$ is below $0.5$, where the last condition was introduced to reduce the chances of being stuck at a local minimum in the optimization procedure.
To avoid an underestimation of MSE due to overfitting, in this optimization process we only use half of the pairs $(n,\bar\gamma_n)$ --- namely, the ones in even positions in the list --- in the set $S_{\mathrm{reg}}$ to estimate the regression coefficients, while we use all of them to calculate the MSE.
This partial splitting strategy provides a simple safeguard against overfitting while preserving sufficient data for stable error estimation.

\begin{table}[ht]
\tbl{Optimal polynomial degrees $D_{\gamma}$, $D_v$, and $D_{v^{\rm\ell}}$ for the estimation of the quantities $\bar\Gamma$, $V$, and $V^{\mathrm\ell}$ respectively, for the different coefficients and weighting schemes described in Section \ref{sec:w_coeff}, based on the optimization procedure described in Section \ref{sssec:degree}.}
{\begin{tabular}{ccc|ccc}
\toprule
\spacer{Coefficient} & Weighting Scheme & \spacer{\spacer{$f(i)$}}  & \spacer{$D_{\gamma}$} & \spacer{$D_v$} & \spacer{$D_{v^{\rm\ell}}$} \\
\midrule
Spearman & additive  & $1/i$          & 3 & 4 & 4\\
         &       & $1/i^2$ & 3 & 2 & 3\\
         &       & $1/(i+1)^2$ & 4 & 3 & 2\\  
         &       & $1/(i+2)^2$ & 4 & 3 & 5\\  

         & multiplicative & $1/i$        & 2 & 3 & 4\\
         &       & $1/i^2$ & 5 & 4 & 3\\
         &       & $1/(i+1)^2$  & 5 & 4 & 6\\  
         &       &  $1/(i+2)^2$ & 4 & 4 & 3\\  

Kendall  & additive  & $1/i$     & 2 & 4 & 3\\
         &       & $1/i^2$ & 4 & 2 & 1\\
         &       & $1/(i+1)^2$ & 5 & 3 & 2\\  
         &       & $1/(i+2)^2$ & 4 & 3 & 5\\  

         & multiplicative & $1/i$    & 2 & 3 & 3\\
         &       & $1/i^2$ & 3 & 1 & 3\\
         &       & $1/(i+1)^2$ & 2 & 1 & 1\\  
         &       & $1/(i+2)^2$ & 2 & 1 & 2\\  
\bottomrule
\end{tabular}}
\label{tab:deg}
\end{table}

The results of this optimization procedure for $D_{\gamma}$ are shown in Table \ref{tab:deg}: we can notice that, depending on the coefficient and the weighting scheme, the optimal degree can vary from 2 to 5.

\section{Estimation of Variance $V$ and Left Variance $V^{\mathrm{\ell}}$}
\label{sec:var_estimation}

Once $\bar\Gamma$ has been either exactly computed (for $n \leq n_{\mathrm{exact}}$) or numerically estimated (for $n > n_{\mathrm{exact}}$), we proceed with the estimation of the variance $V$ and the left variance $V^{\mathrm{\ell}}$.

\subsection{Exact Calculation.}

Similarly to the procedure described in Appendix \ref{sec:gamma_bar_estimate}, the quantities $V$ and $V^{\mathrm{\ell}}$ can be computed exactly for $n \leq n_{\mathrm{exact}}$ by summing over all permutations.
Using the definitions in Eqs.~\ref{eq:variance} and \ref{eq:variance_left} together with the definition of $p(\gamma)$ in Eq.~\ref{eq:notation_x}, we obtain
\begin{eqnarray}
        V &=& \frac{1}{n!}\sum_{\pi\in \mathbb{\Pi}_n}[x(\pi)-\bar\Gamma]^2 ,\\
        \nonumber     V^{\mathrm{\ell}} &=& \frac{1}{n!}\sum_{\pi\in\mathbb{\Pi}_n}[x(\pi)-\bar\Gamma]^2H(\bar\Gamma - x(\pi)),
\end{eqnarray}
where $H(x)$ denotes the Heaviside step function.

\subsection{Numerical Sampling for $V$ and $V^{\mathrm{\ell}}$}

Analogously to the sampling procedure described in Appendix \ref{sec:numerical_sampling}, we estimate $V$ and $V^{\mathrm{\ell}}$ using a sample set $S_{\pi}^n \subset \mathbb{\Pi}_n$ consisting of $N_{\mathrm{samp}}$ permutations independently drawn with uniform probability (allowing repetition).

The estimates of variance and left variance are respectively
\begin{eqnarray}
        v &=& \frac{1}{N_{\mathrm{samp}}}\sum_{\pi\in S_{\pi}^n}[x(\pi)-\bar\gamma]^2 ,\\
        \nonumber     v^{\mathrm{\ell}} &=& \frac{1}{N_{\mathrm{samp}}}\sum_{\pi\in S_{\pi}^n}[x(\pi)-\bar\gamma]^2H(\bar\gamma - x(\pi)),
\end{eqnarray}
where $\bar\gamma$ denotes the regression-based estimate described in Appendix \ref{sec:polynomial_reg}.

\subsection{Polynomial Regression for the Variances}

The polynomial regression procedure for the estimate of $V$ and $V^{\mathrm{\ell}}$ follows the same structure described in Appendix \ref{sec:polynomial_reg} for $\bar\Gamma$, with the following differences.
\begin{itemize}
  \item  The objective function is the standard (unweighted) sum of squared residuals.
  \item The logarithmic transformation $x(n)=1/{\log}(n)$ is adopted only for the multiplicative weighting scheme applied to the weighting function $1/(n + n_0)^2$, as it does not provide satisfactory behavior in the other cases.
\end{itemize}

The results of this optimization procedure are shown in Table \ref{tab:deg}. 
We observe that, depending on the coefficient and weighting scheme, the optimal polynomial degree ranges from 1 to 6.

\section{Distributional Parameters Estimation Results}
\label{sec:estimation_results}

As discussed in Section \ref{sec:flat_variance_ratio}, symmetric ranking coefficients such as
the standard Spearman $\rho$ and Kendall $\tau$ are already in standardized form and therefore require no shift.

\begin{itemize}
      \item The coefficients used to estimate $\bar\Gamma$, $V$, and $V^{\mathrm{\ell}}$ are denoted respectively by $y_a$, $z_a$, and $w_a$.
      \item These coefficients must be applied to the transformed ranking length $x(n)$ defined in Eqs.~\ref{eq:n_to_x} and \ref{eq:n_to_x_log}, where the latter is used in all cases corresponding to the value \emph{True} for the \emph{${\it is\_log}$} field.
\end{itemize}

\begin{table}[ht]
\tbl{Results for Spearman coefficient with additive and multiplicative weighting scheme.}
{\begin{tabular}{c|cccc|cccc}
\toprule
 &\multicolumn{4}{c|}{Additive weighting scheme} &  \multicolumn{4}{c}{Multiplicative weighting scheme} \\
 $f(i)$  & $1/i$ & $1/i^2$& $1/(i+1)^2$ & $1/(i+2)^2$ & $1/i$ & $1/i^2$& $1/(i+1)^2$ & $1/(i+2)^2$ \\
\midrule
${\bar\Gamma}_{3}$ & -0.0351391 & -0.12458 & -0.0542795 & -0.0302762 & -0.0351391 & -0.12458 & -0.0542795 & -0.0302762 \\ 
${\bar\Gamma}_{4}$ & -0.055027 & -0.1860625 & -0.0908478 & -0.0542042& -0.0351391 & -0.12458 & -0.0542795 & -0.0302762  \\  
${\bar\Gamma}_{5}$ & -0.0699207 & -0.2311256 & -0.1214359 & -0.0758427 & -0.1105363 & -0.3333887 & -0.1816873 & -0.1138495 \\ 
${\bar\Gamma}_{6}$ & -0.0820016 & -0.2667351 & -0.1480683 & -0.0958261 & -0.115858 & -0.3571699 & -0.1976062 & -0.125526 \\ 
${\bar\Gamma}_{7}$ & -0.0921957 & -0.2959716 & -0.1716402 & -0.1143618  & -0.1190921 & -0.3731397 & -0.2094618 & -0.1346007 \\ 
${\bar\Gamma}_{8}$ & -0.1010127 & -0.3205778 & -0.1927137 & -0.1315846 & -0.1211498 & -0.3844621 & -0.218712 & -0.1419613 \\ 
${\bar\Gamma}_{9}$ & -0.1087727 & -0.3416628 & -0.211696 & -0.1476116 & -0.1224877 & -0.3927709 & -0.2261579 & -0.1481048 \\ 
${\bar\Gamma}_{10}$ & -0.1156932 & -0.3599846 & -0.2289015 & -0.1625505 & -0.1233591 & -0.3990102 & -0.2322883 & -0.1533393 \\
$n_{\mathrm{max}}^{\gamma}$  & 40000 & $\infty$ & $\infty$ & $\infty$ & 40000 & 40000 & 40000 & 40000 \\  
${\mathrm{is\_log}}^{\gamma}$  & True & False & False & False  & True & True & True & True \\  
$y_0$  & -0.601437 & -0.661424 & -0.62854 & -0.616849 & -0.019307 & 0.038146 & 0.090947 & 0.306574 \\
$y_1$  & 2.341654 & 6.062951 & 11.59787 & 15.79627 & -0.595729 & -0.080508 & -1.017228 & -6.368495 \\ 
$y_2$  & -3.801763 & -55.79642 & -214.1718 & -330.0138 & 0.837962 & -33.42217 & -23.15904 & 27.40441 \\ 
$y_3$  & 2.248584 & 261.0996 & 2477.381 & 3905.447  & -        & 185.8295 & 159.0246 & -48.33406 \\ 
$y_4$  & -        & -        & -11180.82 & -17614.87   & -        & -388.7535 & -367.7098 & 30.79281 \\
$y_5$ & - & - & - & - & -        & 292.1339 & 294.2084 & -        \\ 
${V}_{3}$ & 0.5181185 & 0.5601135 & 0.5216072 & 0.5113228 & 0.5333613 & 0.6345766 & 0.549567 & 0.5222276 \\ 
${V}_{4}$ & 0.3670496 & 0.4168125 & 0.3755816 & 0.3586697 & 0.3820184 & 0.4515291 & 0.3917036 & 0.3706945 \\
${V}_{5}$ & 0.2900554 & 0.3414761 & 0.3011874 & 0.2824563 & 0.3065119 & 0.3545254 & 0.3137249 & 0.2957158 \\ 
${V}_{6}$ & 0.2426134 & 0.2930327 & 0.2550984 & 0.2361664 & 0.2604369 & 0.2958008 & 0.2665942 & 0.2505371 \\  
${V}_{7}$ & 0.2101364 & 0.2586132 & 0.2232419 & 0.2047564 & 0.2290399 & 0.2564279 & 0.2346528 & 0.2201062 \\ 
${V}_{8}$ & 0.1863572 & 0.2326099 & 0.1996417 & 0.1818611  & 0.2060967 & 0.2281286 & 0.2113929 & 0.1980887 \\ 
${V}_{9}$ & 0.1681119 & 0.212133 & 0.181306 & 0.164317 & 0.1884992 & 0.2067728 & 0.1935963 & 0.1813411 \\ 
${V}_{10}$ & 0.1536211 & 0.1955189 & 0.1665609 & 0.15037 & 0.1745125 & 0.1900649 & 0.1794765 & 0.1681206 \\  
$n_{\mathrm{max}}^{v}$  & $\infty$ & $\infty$ & $\infty$ & $\infty$ & $\infty$ & 40000 & 40000 & 40000 \\ 
${\mathrm{is\_log}}^{v}$  & False & False & False & False & False & True & True & True \\  
$z_0$  & 0.004487 & 0.026646 & 0.012705 & 0.008131  & 0.012538 & -0.033754 & 0.010193 & 0.018235 \\  
$z_1$  & 2.988307 & 1.948141 & 1.973444 & 1.883465 & 3.697446 & 0.688005 & -0.298537 & -0.50893 \\ 
$z_2$  & -57.8245 & -2.76501 & -7.432244 & -8.196073   & -49.50208 & -2.742296 & 3.34995 & 4.781896 \\ 
$z_3$  & 816.9964 & -        & 33.06051 & 36.222  & 300.3544 & 8.618715 & -6.157762 & -10.18314 \\ 
$z_4$  & -3957.798 & -        & -        & -      & -        & -7.441726 & 4.85662 & 8.548069 \\ 
${V^{{\rm\ell}}}_{3}$ & 0.245896 & 0.2255373 & 0.2393852 & 0.2441574 & 0.2260603 & 0.2183891 & 0.2123318 & 0.2236318 \\ 
${V^{{\rm\ell}}}_{4}$ & 0.1685735 & 0.160236 & 0.1641092 & 0.1647886 & 0.160053 & 0.1354465 & 0.1491632 & 0.1553757 \\  
${V^{{\rm\ell}}}_{5}$ & 0.1322988 & 0.1287807 & 0.1295744 & 0.1281768 & 0.1312279 & 0.1077399 & 0.1234245 & 0.1266716 \\ 
${V^{{\rm\ell}}}_{6}$ & 0.1105955 & 0.1092839 & 0.1086981 & 0.1065389 & 0.1130839 & 0.092622 & 0.1058792 & 0.1082567 \\  
${V^{{\rm\ell}}}_{7}$ & 0.0957123 & 0.0957021 & 0.0943793 & 0.0917853 & 0.1001576 & 0.0809675 & 0.0931088 & 0.0952586 \\  
${V^{{\rm\ell}}}_{8}$ & 0.0848307 & 0.0856157 & 0.0838575 & 0.0810663 & 0.090589 & 0.0725216 & 0.0840994 & 0.0858235 \\ 
${V^{{\rm\ell}}}_{9}$ & 0.0764926 & 0.0777563 & 0.07573 & 0.0728762 & 0.0832859 & 0.0666184 & 0.0775753 & 0.0789453 \\ 
${V^{{\rm\ell}}}_{10}$ & 0.0698778 & 0.0714465 & 0.0692395 & 0.0663905 & 0.0775167 & 0.0622902 & 0.0726842 & 0.0736614 \\ 
$n_{\mathrm{max}}^{v^{{\rm\ell}}}$  & $\infty$ & $\infty$ & $\infty$ & $\infty$ & $\infty$ & 40000 & 40000 & 40000 \\ 
${\mathrm{is\_log}}^{v^{{\rm\ell}}}$  & False & False & False & False & False & True & True & True \\ 
$w_0$  & 0.00208 & 0.01059 & 0.00568 & 0.003742 & 0.004903 & -0.013756 & -0.00869 & -0.005547 \\ 
$w_1$  & 1.410178 & 0.74149 & 0.727979 & 0.728583 & 2.397079 & 0.304449 & 0.120127 & 0.041578 \\  
$w_2$  & -29.80098 & -2.920581 & -0.952968 & -1.128635 & -68.24124 & -0.746029 & 0.157677 & 0.52811 \\ 
$w_3$  & 441.8369 & 16.77604 & -        & -        & 1006.321 & 1.043273 & -        & -0.477712 \\
$w_4$  & -2221.862 & -        & -        & -    & -5002.647 & -        & -        & -  
 \\ \bottomrule 
\end{tabular}}
\label{tab:result_spearman}
\end{table}

\begin{table}[ht]
\tbl{Results for Kendall coefficient with additive and multiplicative weighting scheme.}
{\begin{tabular}{c|cccc|cccc}

\toprule
 &\multicolumn{4}{c|}{Additive weighting scheme} &  \multicolumn{4}{c}{Multiplicative weighting scheme} \\
 $f(i)$  & $1/i$ & $1/i^2$& $1/(i+1)^2$ & $1/(i+2)^2$ & $1/i$ & $1/i^2$& $1/(i+1)^2$ & $1/(i+2)^2$ \\
\midrule

${\bar\Gamma}_{3}$ & -0.0428591 & -0.1392512 & -0.0643976 & -0.0366118 & -0.0745921 & -0.185488 & -0.1028431 & -0.0637739 \\ 
${\bar\Gamma}_{4}$ & -0.0618355 & -0.2074294 & -0.1003907 & -0.0593039 & -0.0992074 & -0.26514 & -0.1459207 & -0.0905944 \\ 
${\bar\Gamma}_{5}$ & -0.0741663 & -0.2522021 & -0.1274014 & -0.0778002 & -0.1114205 & -0.3109147 & -0.1722248 & -0.1076588 \\  
${\bar\Gamma}_{6}$ & -0.0834441 & -0.2852926 & -0.1496204 & -0.0940042 & -0.1188077 & -0.3414211 & -0.1911483 & -0.1204973 \\  
${\bar\Gamma}_{7}$ & -0.0909502 & -0.311327 & -0.168656 & -0.1085925 & -0.1238325 & -0.363651 & -0.2059938 & -0.1309927 \\
${\bar\Gamma}_{8}$ & -0.0972845 & -0.3326212 & -0.1853389 & -0.1219045 & -0.1275243 & -0.3808363 & -0.218257 & -0.1399837 \\ 
${\bar\Gamma}_{9}$ & -0.1027778 & -0.3505098 & -0.2001783 & -0.1341517 & -0.1303874 & -0.394686 & -0.2287325 & -0.1479128 \\  
${\bar\Gamma}_{10}$ & -0.1076337 & -0.3658366 & -0.213521 & -0.1454849 & -0.1326976 & -0.406195 & -0.2378916 & -0.1550424 \\ 
$n_{\mathrm{max}}^{\gamma}$  & 3000 & $\infty$ & $\infty$ & $\infty$ & 3000 & 3000 & 3000 & 3000 \\ 
${\mathrm{is\_log}}^{\gamma}$  & True & False & False & False & True & True & True & True \\
$y_0$  & -0.452135 & -0.61982 & -0.544036 & -0.521584 & -0.235206 & -1.012771 & -0.889377 & -0.877253 \\ 
$y_1$  & 1.49774 & 6.387459 & 11.26601 & 14.80992 & 0.434552 & 3.502608 & 2.860412 & 3.401867 \\ 
$y_2$  & -1.648071 & -119.0659 & -286.6257 & -334.0915 & -0.4637 & -8.290404 & -3.212251 & -4.108049 \\ 
$y_3$  & -        & 1550.544 & 5118.082 & 3972.444  & -        & 8.039727 & -        & -        \\ 
$y_4$  & -        & -7595.937 & -47580.03 & -17689.92  & -        & -  & -        & -        \\ 
$y_5$  & -        & -        & 171722.3 & - & -        & -  & -        & -        \\ 

${V}_{3}$ & 0.4643206 & 0.5865483 & 0.4861446 & 0.4512905 & 0.5406241 & 0.7021316 & 0.5839694 & 0.5203415 \\ 
${V}_{4}$ & 0.3005303 & 0.4268329 & 0.3277733 & 0.2915696 & 0.3799759 & 0.5494523 & 0.4344145 & 0.3692972 \\
${V}_{5}$ & 0.2237503 & 0.3422718 & 0.2530008 & 0.2187594 & 0.3003558 & 0.4608244 & 0.3601742 & 0.2990175 \\ 
${V}_{6}$ & 0.1796905 & 0.2894809 & 0.2096939 & 0.1777569 & 0.2531364 & 0.4037092 & 0.316453 & 0.2594437 \\
${V}_{7}$ & 0.1512146 & 0.2530892 & 0.1813923 & 0.1515961 & 0.2219125 & 0.3640513 & 0.2878464 & 0.2344817 \\ 
${V}_{8}$ & 0.1313125 & 0.2263027 & 0.1613661 & 0.1334668 & 0.1997066 & 0.3349671 & 0.2677464 & 0.2175054 \\
${V}_{9}$ & 0.1166122 & 0.2056522 & 0.1463784 & 0.120144 & 0.1830722 & 0.3127349 & 0.2528784 & 0.2053251 \\ 
${V}_{10}$ & 0.1052991 & 0.189177 & 0.1346872 & 0.1099161 & 0.1701178 & 0.2951823 & 0.2414456 & 0.1962291 \\ 
$n_{\mathrm{max}}^{v}$  & $\infty$ & $\infty$ & $\infty$ & $\infty$ & $\infty$ & 3000 & 3000 & 3000 \\ 
${\mathrm{is\_log}}^{v}$  & False & False & False & False & False & True & True & True \\ 
$z_0$  & 0.005522 & 0.020643 & 0.015561 & 0.010423 & 0.036587 & 0.025408 & 0.074257 & 0.115488 \\
$z_1$  & 1.761004 & 2.178677 & 1.879833 & 1.803538 & 2.836327 & 0.583835 & 0.368796 & 0.154672 \\ 
$z_2$  & -32.87796 & -5.440805 & -13.76867 & -19.08445 & -36.09014 & -        & -        & -        \\ 
$z_3$  & 502.6566 & -        & 68.77372 & 113.1593 & 218.8208 & -        & -        & -        \\ 
$z_4$  & -2569.777 & -        & -        & -        & - & -        & -        & -        \\ 

${V^{{\rm\ell}}}_{3}$ & 0.2018534 & 0.212235 & 0.1988788 & 0.1991994 & 0.2068706 & 0.2443723 & 0.215398 & 0.202704 \\
${V^{{\rm\ell}}}_{4}$ & 0.128829 & 0.1330324 & 0.1286455 & 0.1264909 & 0.1345549 & 0.1595142 & 0.1437786 & 0.1350048 \\
${V^{{\rm\ell}}}_{5}$ & 0.0952204 & 0.0997918 & 0.0973095 & 0.0934387 & 0.1052886 & 0.1236619 & 0.1146279 & 0.1073594 \\ 
${V^{{\rm\ell}}}_{6}$ & 0.0760413 & 0.0815773 & 0.0790937 & 0.0747621 & 0.087606 & 0.1032788 & 0.0974049 & 0.0906875 \\ 
${V^{{\rm\ell}}}_{7}$ & 0.0637416 & 0.0697496 & 0.0672671 & 0.0629253 & 0.0754118 & 0.0902007 & 0.0862253 & 0.0797315 \\ 
${V^{{\rm\ell}}}_{8}$ & 0.0552042 & 0.0613258 & 0.0590139 & 0.0547863 & 0.0668054 & 0.0811217 & 0.0785152 & 0.0722813 \\ 
${V^{{\rm\ell}}}_{9}$ & 0.048932 & 0.0549977 & 0.0529286 & 0.0488417 & 0.0604781 & 0.0744639 & 0.0728941 & 0.0669478 \\ 
${V^{{\rm\ell}}}_{10}$ & 0.0441259 & 0.0500832 & 0.0482469 & 0.0443035 & 0.0556269 & 0.0693738 & 0.0686167 & 0.0629382 \\ 
$n_{\mathrm{max}}^{v^{{\rm\ell}}}$  & $\infty$ & $\infty$ & $\infty$ & $\infty$ & $\infty$ & 3000 & 3000 & 3000 \\  
${\mathrm{is\_log}}^{v^{{\rm\ell}}}$  & False & False & False & False & False & True & True & True \\ 
$w_0$  & 0.002905 & 0.010087 & 0.00762 & 0.005096 & 0.009314 & -0.011199 & 0.011014 & 0.035322 \\ 
$w_1$  & 0.573667 & 0.399278 & 0.455188 & 0.463824 & 0.852271 & 0.307276 & 0.121451 & -0.038073 \\ 
$w_2$  & -3.503062 & -        & -0.628606 & -0.839183 & -8.693876 & -0.766444 & -        & 0.223556 \\ 
$w_3$  & 17.85528 & -        & -        & -  & 48.29288 & 1.137351 & -        & -  \\
\bottomrule 
\end{tabular}}
\label{tab:result_kendall_add}
\end{table}

The results reported in this section were obtained using the code available at \url{https://github.com/plombardML/ranking_correlation};
 random sampling was implemented through the pseudo-random number generator included in the \emph{numpy.random} Python library.
 Computations were performed on a machine equipped with an Intel Core i7 processor ($2.80\, {\mathrm{GHz}}$) and $32\,{\mathrm{GB}}$ of RAM.

The repository also includes a Python function \emph{standard\_gamma\_calc()} that implements the complete procedure to estimate $g(x)$ described in Appendix \ref{sec:g0_determination_algorithm}.

We do not report results for $n=2$, as no standardization is required in this case. In fact, with only one possible pairwise exchange, the symmetry described in Section \ref{sec:EV_standard} holds also for weighted coefficients, yielding $\bar\Gamma=0$, $V=1$, and $V^{\rm\ell}=1/2$.

\end{document}